\documentclass[sigconf]{acmart}

\AtBeginDocument{%
  }

\setcopyright{acmlicensed}
\copyrightyear{2018}
\acmYear{2018}
\acmDOI{XXXXXXX.XXXXXXX}
\acmConference[Conference acronym 'XX]{Make sure to enter the correct
  conference title from your rights confirmation email}{June 03--05,
  2018}{Woodstock, NY}
\acmISBN{978-1-4503-XXXX-X/2018/06}

\usepackage{graphicx} 

\usepackage{booktabs} 
\usepackage{adjustbox} 
\usepackage{array}    
\usepackage{multirow}
\usepackage{subcaption}
\usepackage{hyperref}

\usepackage{booktabs}
\usepackage{array}
\usepackage{xcolor}
\usepackage{siunitx}
\definecolor{better}{RGB}{0,100,0}
\definecolor{worse}{RGB}{180,0,0}
\usepackage{xcolor}
\usepackage{enumitem}
\usepackage{amsmath}

\usepackage{caption}
\usepackage{blindtext}
\usepackage{tcolorbox}
\usepackage[final]{pdfpages}
\usepackage{lipsum,multicol}
\usepackage{xcolor}
\usepackage{tikz}
\usepackage{listings}
\usepackage{enumitem}
\usepackage{hyperref}
\usepackage{amsfonts}
\usepackage{wrapfig}
\usepackage{subcaption} 
\usepackage{adjustbox}
\usepackage{colortbl}
\usepackage{fancybox}
\usepackage{multirow}
\usepackage[normalem]{ulem}
\useunder{\uline}{\ul}{}
\usepackage{enumitem}
\usepackage{subfiles}

\lstdefinestyle{javastyle}{
    language=Java,
    basicstyle=\ttfamily\small,
    keywordstyle=\color{blue}\bfseries,
    commentstyle=\color{gray},
    stringstyle=\color{red},
    numbers=left,
    numberstyle=\tiny,
    stepnumber=1,
    breaklines=true,
    frame=single
}

\newcommand{\eg}{\textit{e.g.,}\xspace}

\newcommand{\fix}[2]{{\color{black}#2}}

\newtcolorbox{boxK}{
    fontupper = \small,
    sharpish corners, 
    boxrule = 0pt,
    toprule = 0pt, 
}

\newcommand{\llmsforcode}{LLMs4Code\xspace}

\begin{document}

\title{How Quantization Impacts Privacy Risk on LLMs for Code?}

\author{
  Md Nazmul Haque$^{1}$,
  Hua Yang$^{1}$,
  Zhou Yang$^{2}$,
  Bowen Xu$^{1}$
}

\affiliation{%
  \institution{$^{1}$North Carolina State University, $^{2}$University of Alberta}
  \country{}
}

\email{{mhaque4, hyang45, bxu22}@ncsu.edu, zy25@ualberta.ca}

\begin{abstract}
Large language models for code (\llmsforcode) rely heavily on massive training data, including sensitive data, such as cloud service credentials of the projects and personal identifiable information of the developers, raising serious privacy concerns. Membership inference (MI) has recently emerged as an effective tool for assessing privacy risk by identifying whether specific data belong to a model's training set. In parallel, model compression techniques, especially quantization, have gained traction for reducing computational costs and enabling the deployment of large models. However, while quantized models still retain knowledge learned from the original training data, it remains unclear whether quantization affects their ability to retain and expose privacy information. Answering this question is of great importance to understanding privacy risks in real-world deployments.

In this work, we conduct the first empirical study on how quantization influences task performance and privacy risk simultaneously in \llmsforcode. To do this, we implement widely used quantization techniques (static and dynamic) to three representative model families, namely Pythia, CodeGen, and GPT-Neo. Our results demonstrate that quantization has a significant impact on reducing the privacy risk relative to the original model.
We also uncover a positive correlation between task performance and privacy risk, indicating an underlying trade-off. 
Moreover, we reveal the possibility that quantizing larger models could yield better balance than using full-precision small models.
Finally, we demonstrate that these findings generalize across different architectures, model sizes and MI methods, offering practical guidance for safeguarding privacy when deploying compressed \llmsforcode. \end{abstract}

\setcopyright{none} 
\settopmatter{printacmref=false} 
\renewcommand\footnotetextcopyrightpermission[1]{}
\maketitle


\section{Introduction}\label{sec:intro}



The advent of Large Language Models for Code (\llmsforcode) represents a significant milestone in software engineering, specifically engineered to assist in software development by understanding, generating, and manipulating programming code \cite{hou2024large}. 
These models, often trained on vast repositories of source code from GitHub~\cite{gao2020pile, husain2019codesearchnet, allamanis2013mining}, offer substantial productivity gains through features like code completion, test case generation, summarization, bug detection, code translation, etc~\cite {li2023starcoder, lu2021codexglue, nijkamp2023codegen}. 

However, their dependence on vast amounts of training data raises significant privacy concerns. On the one hand, one critical concern is the potential violation of intellectual property rights through the misuse of code governed by restrictive licenses. In particular, malicious developers might train \llmsforcode on license-restricted code, yet remain undetected due to the lack of effective auditing mechanisms. To mitigate such threats, recent studies explored Membership Inference (MI) methods as a detection mechanism~\cite{wan2024does, zhou2024gotcha}. MI reveals whether a specific data point, such as a piece of code, was included in the \llmsforcode training dataset. Prior research has demonstrated that MI can effectively identify unauthorized usage of license-restricted code, highlighting its potential for auditing and enforcing compliance in large-scale model training~\cite{zhou2024gotcha}. On the other hand, a successful MI poses a critical threat: it could reveal that proprietary algorithms, confidential business logic, or sensitive data embedded within code comments or strings (e.g., credentials, API keys) were part of the training corpus, which can serve as an attack vector for a backdoor attack.

Concurrently, the practical deployment of \llmsforcode is hindered by its sheer scale. These models often comprise billions of parameters, demanding substantial computational resources for training and inference, significant memory allocation, and resulting in considerable latency \cite{huang2025pruning}. To address these challenges, model compression techniques—primarily pruning, quantization, and knowledge distillation—have become indispensable tools \cite{huang2025pruning}. These techniques aim to reduce the model's size and computational requirements, thereby facilitating deployment in resource-constrained environments, such as edge devices, or enabling faster and more cost-effective inference in cloud settings \cite{xu2023survey}. Hong et al. found that quantization is currently a more effective technique than others in achieving efficiency \cite{hong2024decoding}.
The quantized version of many well-known \llmsforcode have been released on HuggingFace, such as CodeLlama-13B~\cite{codellama-awq} and StarCoder2-15B~\cite{starcoder2-gguf}.


While quantized models reduce memory and computational costs, their impact on privacy risk remains an open and critical question. 
Intriguingly, the very act of reducing precision through quantization could inadvertently sharpen the model's output characteristics. This increased determinism, where predictions become more consistent for a given input, might create a more discernible fingerprint of the training data. Attackers could potentially exploit these subtle, yet consistent, output patterns to more easily differentiate between training data and non-training data, thereby increasing MI risks.
In privacy-sensitive applications, such risks are exacerbated. Despite the potential impact of quantization on MI, its effect has not been empirically validated yet. Hence, we propose our central research question (RQ):
\vspace{-1.5mm}
\begin{tcolorbox}
\textit{How does quantization impact privacy risk on \llmsforcode?}
\end{tcolorbox}
\vspace{-1.5mm}

To answer this RQ, we experimented with quantization techniques and evaluated their effects on three well-known families of models, \texttt{Pythia}, \texttt{CodeGen}, and \texttt{GPT-Neo}. Then, we assessed the impact of quantization on both task and MI performance. 
Our results reveal a nuanced trade-off: 8-bit static quantization can preserve task performance while significantly reducing privacy risk compared to the original models, whereas 4-bit quantization can also significantly reduce privacy but at the cost of substantial drops in task performance.
This suggests that while quantization offers potential privacy-preserving benefits, lower bit quantization may compromise task performance. Also, we found scenarios where a lower bit quantization of a larger model yields resilient task performance while improving privacy compared to smaller, original models. The main contributions of this study are as follows:

\begin{itemize}[leftmargin=*]
\item 
We conduct the first empirical study on how quantization affects privacy risk in \llmsforcode. Along with this, we also analyze its effect on task performance. We find that 8-bit quantization maintain comparable task performance with significant privacy improvement, while 4-bit quantization reduce the privacy risk but at a cost of significant task performance drops, especially in smaller models.
We uncover a positive correlation between task performance and MI effectiveness, revealing a fundamental privacy risk-effectiveness trade-off in quantized models.
\item We validate our findings across three \llmsforcode families such as \texttt{Pythia}, \texttt{CodeGen}, and \texttt{GPT-Neo} and demonstrate that the observed trends generalize across architectures and model sizes. 
\item We identify cases where quantized larger models achieve comparable or even smaller model sizes than their full-precision smaller counterparts, while delivering better task performance and lower privacy risks. This highlights strategic quantization as a dual-purpose solution for efficiency and privacy.
\item We provide an actionable guideline for model developers to identify optimal quantization configurations.
\end{itemize}

\noindent\textbf{Data availability}.
We release experimental results and replication package at \url{https://anonymous.4open.science/r/mimir-4746}.

\section{Background}
\subsection{Quantization for \llmsforcode}

There are two main quantization strategies: Quantization-Aware Training (QAT) and Post-Training Quantization (PTQ). QAT~\cite{Esser2020LEARNED} incorporates quantization operations into the training loop, enabling the model to adapt to precision loss. However, due to its high computational and memory requirements, QAT is often impractical for large-scale models. In contrast, PTQ~\cite{Cai_2020_CVPR} performs quantization on a pre-trained model without requiring access to the original training procedure. PTQ typically relies on a calibration dataset to adjust the model’s scaling factors and minimize quantization errors, offering a much lower computational burden. \fix{1.3}{Because of the above advantages, we choose PTQ over QAT as the quantization strategy in our experiment.}

In the context of quantization specifically for \llmsforcode, Wei et al.~\cite{wei2023tpwards} conducted an empirical study demonstrating that applying PTQ to a 16B-parameter model reduced its memory usage by 29\% with negligible performance degradation. To optimize inference speed, Sun et al.~\cite{sun2024when} proposed a Dynamic Inference method that improves efficiency for code completion tasks by selectively skipping layers, achieving an 11.2\% reduction in latency. Extending this line of research, \cite{giagnorio2025quantizing} investigated extreme quantization on LLM4Code models up to 33B parameters. Their study examined advanced PTQ techniques under various calibration settings—including code-specific datasets—and revealed that 4-bit quantization can reduce memory consumption by approximately 70\% on average, without significant loss in code generation quality.
Different from the prior works, our study investigates the impact of quantization uniquely from a privacy perspective. 

\subsection{Membership Inference on \llmsforcode}

MI methods are invented to determine whether a specific data sample was included in the training set of a target model. Hence, MI has been widely used to measure the privacy risk of an LLM.
The earliest formalization of MI was proposed by Shokri et al.~\cite{shokri2017membership}, who leveraged multiple shadow models to approximate the decision boundaries of the target model. Later, Yeom et al.~\cite{yeom2018privacy} showed that the training loss of a model on a given input could effectively estimate its membership status, linking overfitting to privacy leakage. Li et al.~\cite{zheng2021membership} demonstrated that even under constrained conditions—where only the final predicted label is exposed—MI remain feasible. Carlini et al.~\cite{carlini2021extracting} introduced a novel strategy based on the zlib compression size of an input sample as a proxy for difficulty, setting instance-specific thresholds to infer membership.

Recent reference-based MI methods~\cite{fu2024membership, ye2022enhanced, carlini2023quantifying} have shown improved effectiveness by calibrating model outputs with auxiliary reference models. However, these methods are typically expensive and impractical in real-world scenarios due to the requirement of training and maintaining high-quality reference models that closely align with the target architecture.

While MI has been extensively studied in the domains of image and text classification, its application to code-related models and data remains limited. Yang et al.~\cite{zhou2024gotcha} proposed an MI approach for auto-regressive code pre-trained language models (CPLM), confirming its effectiveness on code generation tasks. More recently, Zhang et al.~\cite{zhang-etal-2024-code} introduced a comprehensive framework for Code Membership Inference that integrates signal extraction from pre-training tasks, calibration via hard-to-learn samples, and weighted inference strategies. 
\vspace{-3mm}
\subsection{Preliminary Results}
\label{sec:prelim}

The most relevant prior work is by Wei et al.~\cite{wei2023tpwards}. While they explored the effects of quantization, their study has several limitations. First, it did not consider the code completion task, which is widely used in practice and has been shown to be particularly vulnerable to privacy leakage~\cite{al2024traces, duan2024membership}. Second, their analysis did not consider popularly used quantization techniques \cite{10.5555/3600270.3602468, dettmers2023qlora, dettmers2022optimizers}; rather, they focused solely on the static quantization technique proposed by themselves.


To address the above limitations, we conducted our preliminary experiment. To address the first limitation, we focus on the models from the Pythia model family for two main reasons. First, Pythia models are widely adopted for code completion~\cite{al2024traces, duan2024membership}. Second, there are multiple variants of Pythia models in different scales, which enable a controlled experiment to observe the impact of model sizes. To address the second limitation, we evaluate both static and dynamic quantization techniques: for static quantization, we consider 8-bit and 4-bit precision levels, while for dynamic quantization, we include only the 8-bit variant, as 4-bit dynamic quantization is not currently supported in PyTorch.

\begin{figure}
    \centering
    \includegraphics[width=0.4\textwidth]{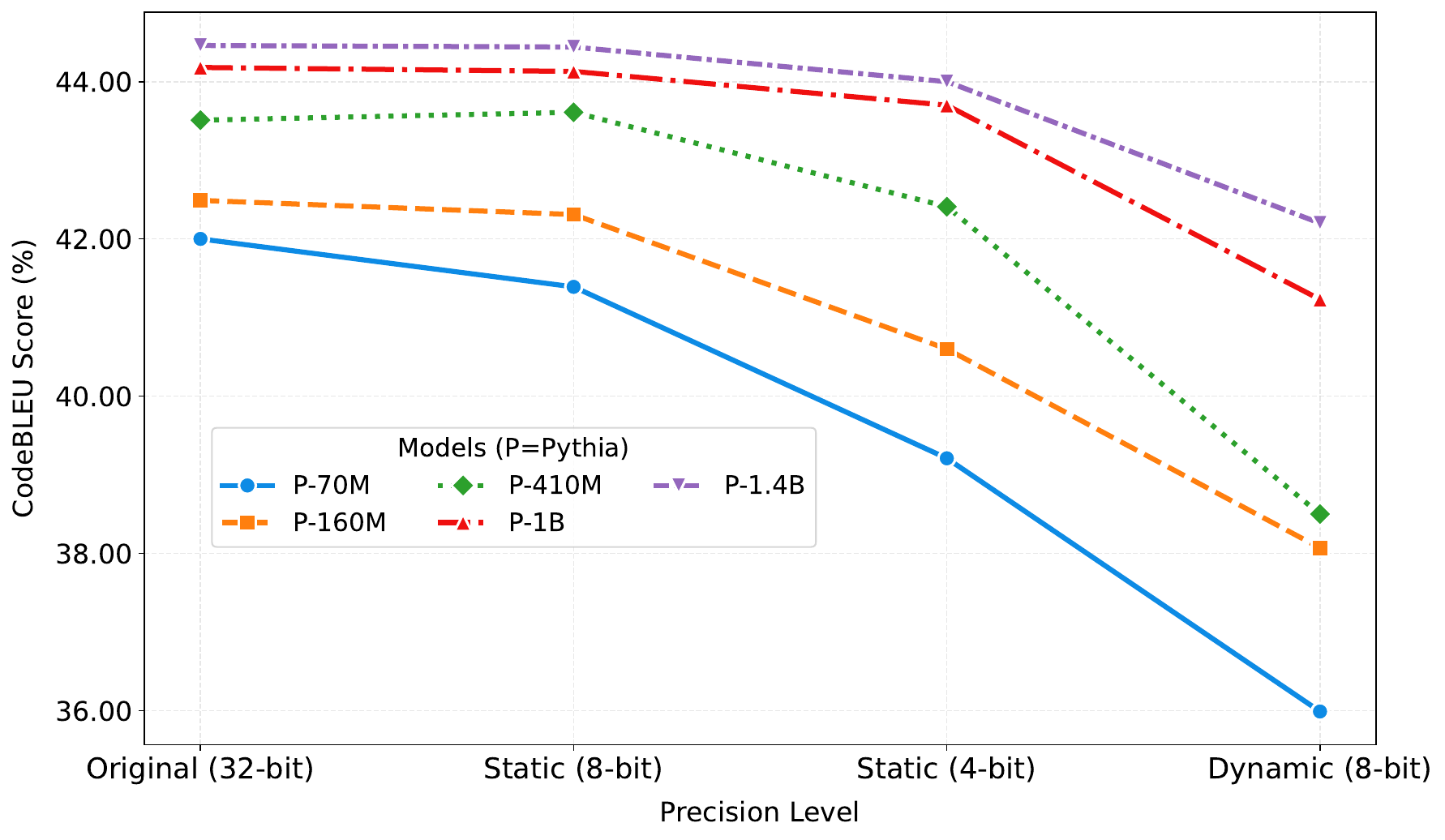}
    \vspace{-3mm}
    \caption{Task Performance (CodeBLEU) of the original and quantized Pythia models. Lines represent different Pythia models distinguished by parameter sizes: Pythia-70M (\textcolor[HTML]{0d8be5}{blue}), Pythia-160M (\textcolor[HTML]{ff7f0e}{orange}), Pythia-410M (\textcolor[HTML]{2ca02c}{green}), Pythia-1B (\textcolor[HTML]{ef1010}{red}), and Pythia-1.4B (\textcolor[HTML]{9467bd}{purple}).}
    \vspace{-5mm}
    \label{fig:staticVSdyn}
\end{figure}

We focus on measuring the impact on code completion task performance in our preliminary experiment. We use CodeBLEU score~\cite{ren2020codebleu} across five Pythia models under three precision settings: the Original model with full precision (32-bit), 8-bit quantization, and 4-bit quantization. Results are summarized in \autoref{fig:staticVSdyn} and present all the detailed numbers in our replication package. To assess statistical significance for task performance, we apply the Wilcoxon signed-rank test \cite{wilcoxon1945individual} under a 95\% confidence interval.


\textbf{From original to 8-bit static quantization.} From the \autoref{fig:staticVSdyn}, it is shown that for all the Pythia models, 8-bit static quantization maintains comparable performance to the original. The relative performance drop ranges from 0.04\% to 1.45\%. For example, Pythia-70M drops slightly from 42.00 to 41.39 (-1.44\%), and Pythia-1.4B drops negligibly from 44.46 to 44.44 (-0.04\%). Exceptionally, we observe two smaller models (i.e., Pythia-70M and 160M) experience a statistically significant drop in performance, medium and larger models show no significant degradation. This trend suggests that static quantization is highly effective for the code completion task.

\textbf{From original to 4-bit static quantization.} Differently, 4-bit static quantization leads to more noticeable task performance drops ($p<0.05$) as compare with the original one, especially in smaller models. For instance, Pythia-70M experiences a performance reduction of 6.64\% (from 42.00 to 39.21), while larger models such as Pythia-1.4B show only minor declines (1.03\%). This indicates that smaller models are more sensitive to aggressive quantization. On average, 4-bit quantization causes about 3.15\% performance drop.


\textbf{From original to 8-bit dynamic quantization.} Dynamic quantization drastically reduces CodeBLEU scores, with relative performance drops ranging from 5.08\% to 14.31\%, and thus exhibits worse task performance than the original full-precision models. The difference between original and dynamic quantization is statistically significant. For example, Pythia-70M sees the largest drop from 42.00 to 36.00 (-14.31\%), while even the largest model, Pythia-1.4B, drops noticeably from 44.46 to 42.21 (-5.08\%). The degradation is consistent across all model sizes—Pythia-160M declines by 10.40\%, Pythia-410M by 11.51\%, and Pythia-1B by 6.67\%.


Overall, our preliminary results show that \textbf{static quantization achieves superior performance in the code completion task compared to dynamic quantization}.
On average, 8-bit static quantization outperforms dynamic quantization by 10.35\%.
Furthermore,
even 4-bit static quantization outperforms 8-bit dynamic quantization by 7.20\%. 
We also find that 8-bit static quantization maintains task performance across model sizes, while 4-bit leads to noticeable degradation, especially in smaller models.
Given the consistently negative impact of dynamic quantization on task performance, we exclude it from our main experiments.


\begin{boxK}
\vspace{-2mm}
\textit{\textbf{Finding:}} 8-bit and even 4-bit static quantization outperform 8-bit dynamic quantization in code completion task performance by a large margin. Quantizing models to 8-bit precision has minimal impact on task performance (CodeBLEU), with changes typically under 1.4\%, whereas 4-bit quantization consistently reduces CodeBLEU scores, especially in smaller models like \texttt{Pythia-70M}, which shows a drop of up to 6.64\%.
\vspace{-2mm}
\end{boxK}



\section{Research Questions}\label{sec:RQ}

Motivated by our preliminary experiment, we aim to answer the following three leading research questions. 

\textbf{RQ1:} \label{rq:miaPerformance} How do different precision levels of quantization impact privacy risk of \llmsforcode? 

\noindent
\textbf{Motivation:} Quantization is increasingly applied to reduce memory and computation costs for deploying LLMs. However, its impact on privacy-related risks remain underexplored. Since quantization alters internal model representations, it could reduce or amplify signals that MI methods exploit. Without a clear understanding of how precision level affect MI performance, practitioners risk deploying models that are more efficient yet privacy-preserving. RQ1 aims to fill this critical gap by empirically analyzing how different quantization levels influence susceptibility to MIs across multiple MI methods and model scales.


\textbf{RQ2:} \label{rq:correlation} Is there a correlation between task performance and privacy risk of quantized \llmsforcode? 

\noindent
\textbf{Motivation}: 
This question is motivated by the critical trade-offs often faced in deploying compressed models—whether changes in task performance inherently relates to the privacy risks (measured by MI effectiveness). Understanding this relationship is essential for developing quantization strategies that strike a decent balance between performance and privacy, guiding practitioners in making informed deployment decisions.


\textbf{RQ3:} \label{rq:generalizibility} Are our findings generalizable to other models?

\noindent
\textbf{Motivation}
To ensure the robustness and applicability of our findings, it is important to examine whether the trends of RQ1 and RQ2 hold across different model architectures and sizes of . Without such validation, the results risk may be model-specific, limiting their broader relevance to the machine learning and software engineering communities. By evaluating other widely used language models, such as CodeGen and GPT-Neo, across varying parameter scales, we aim to assess the generalizability of our observations and provide stronger empirical foundations for informed model deployment in privacy-sensitive applications.




\section{Study Design}
\begin{figure}
    \centering
    \includegraphics[width=0.45\textwidth]{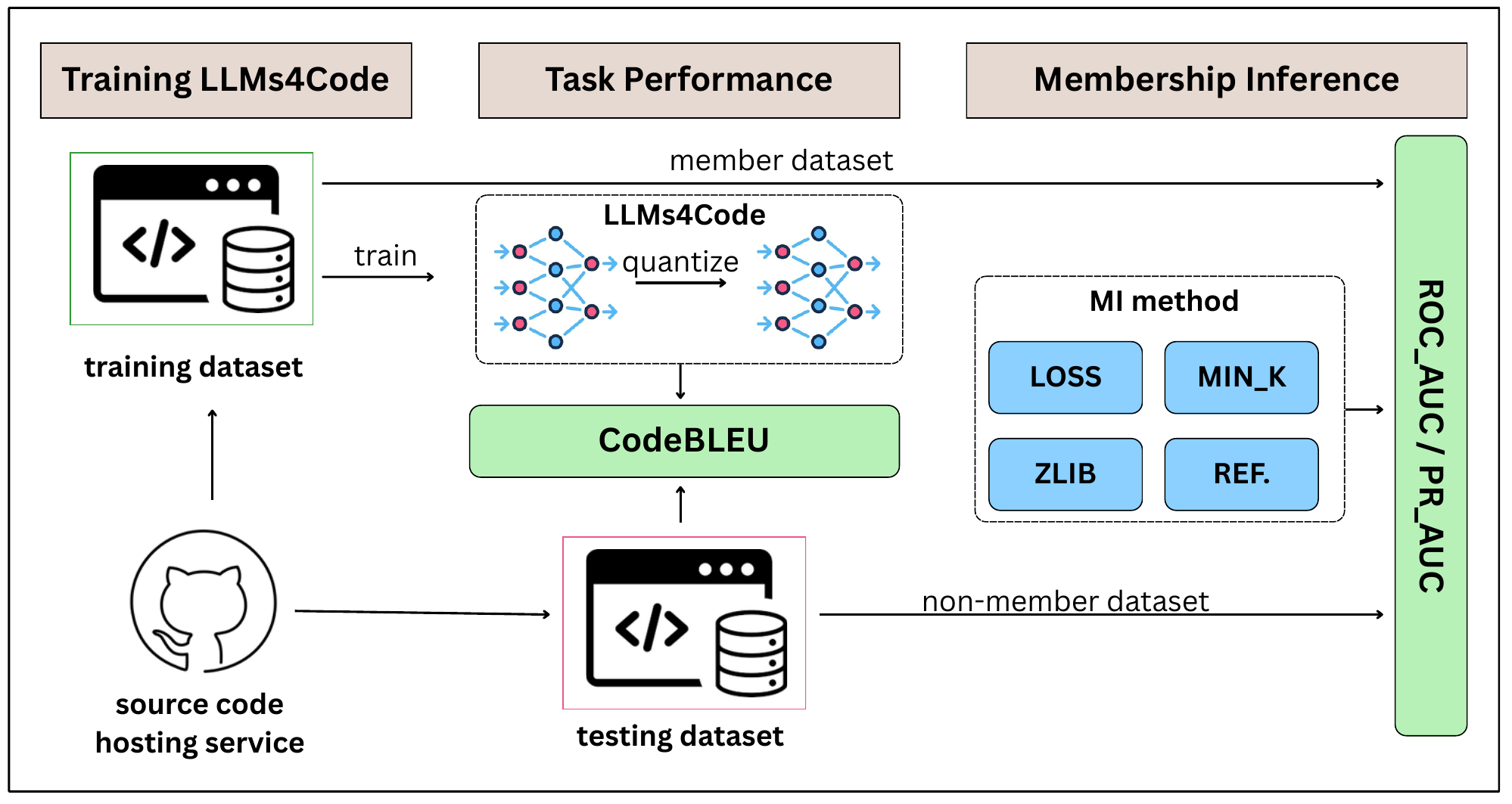}
    \vspace{-3mm}
    \caption{An overview of the study procedure.}
    \label{fig:studyDesign}
    \vspace{-3mm}
\end{figure}

As shown in Figure \ref{fig:studyDesign}, we provide an overview of our study design to answer our core research question described in Section~\ref{sec:intro}. Our methodology involves assessing original and quantized models, analyzing their effectiveness via task performance metrics, and their privacy risks via MI methods. 

\subsection{Dataset Composition}
To assess the privacy risk via MI methods, it requires the information of training data for \llmsforcode. The training data is considered as \textit{member data} to MIs. Based on this, we also need a \textit{non-member} dataset, i.e., a set of data that shares no overlap with the training data. The non-member data could come from either the test set or alternative external sources.

Following the previous study~\cite{duan2024membership}, we utilized a popular dataset derived from the Pile dataset \cite{biderman2022datasheet, gao2020pile}. While the Pile test set is decontaminated at the document level to prevent data leakage, Duan et al.~\cite{duan2024membership} recommended an additional deduplication step, following the methodology of Groeneveld et al. \cite{groeneveldbff}, to further ensure a clean separation between member and non-member samples.
This deduplication process yields a robust evaluation set of 1,000 member and 1,000 non-member samples specifically curated for benchmarking. To ensure experimental consistency and reliability, we used this benchmark dataset as the input for evaluating both task performance and privacy risk across the selected \llmsforcode.


Note that, due to resource limitations, this study did not involve training or fine-tuning \llmsforcode. Instead, we focused on selecting pre-trained models that were already trained on the above known benchmark trained dataset with clear documentation. This approach allows for a meaningful and resource-efficient evaluation of task performance and privacy risk across models of varying sizes and quantization levels.

\subsection{Target Models}
To systematically evaluate both task performance and privacy risk on original and quantized \llmsforcode, we focus on left-to-right autoregressive models available on Hugging Face, \fix{2.4}{ as these are widely used for code completion tasks and have privacy issues \cite{wan2024does, zhou2024gotcha}.}
Our selection prioritizes the models that are specifically trained on the Pile code dataset to ensure the correctness of member and non-member data setting. As a result, we identify three representative families of \llmsforcode—\texttt{CodeGen, GPT-Neo}, and \texttt{Pythia}—each comprising models with varying parameter sizes to ensure diversity in architecture and scale. These model families were selected due to their prevalence in recent research \cite{zhou2024gotcha, duan2024membership} and the availability of multiple parameter configurations. We refer to the selected models as \textit{target models} throughout our evaluation.

However, the well known models, such as DeepMind's AlphaCode, Google’s Gemini Code models (e.g., PaLM 2 for Code), Meta’s CodeLlama, and OpenAI’s ChatGPT and Codex, do not disclose detailed information about their training datasets. This lack of transparency poses a significant barrier to systematically assessing privacy risks.
To ensure fairness and reproducibility in our evaluation, we exclude these models from our analysis and instead focus on publicly available models with well-documented training data.

To assess the impact of model compression on privacy risk, we employ popular quantization techniques. Specifically, we apply both static and dynamic quantization methods. For static quantization, we use the BitsAndBytes library to compress models into 4-bit and 8-bit representations \cite{dettmers2023qlora}. For dynamic quantization, we utilize PyTorch's 8-bit dynamic quantization approach\cite{pytorch-dynamic-quant}; note that 4-bit dynamic quantization is not supported in the current PyTorch implementation (version 3.9).

\subsection{Task Performance Assessment}
To assess the effectiveness of the selected \llmsforcode, we focus on the code completion task, \fix{2.2, 2.3} {specifically at the token level, where the models are evaluated based on their ability to predict the next token given the preceding context. We focus on code completion because it is widely used in development environments and has been identified as a task with high privacy risks, particularly regarding memorization of sensitive content \cite{huang2024your, carlini2021extracting}.} We adopt CodeBLEU \cite{ren2020codebleu}, a widely recognized metric specifically designed for evaluating code generation quality. CodeBLEU extends the traditional BLEU metric \cite{papineni2002bleu} by incorporating code-specific features, such as abstract syntax tree (AST) match, data flow alignment, and code structure similarity, providing a more robust evaluation of the generated code's syntactic and semantic correctness. The CodeBLEU formulation is presented in \autoref{eq:codebleu}.
\begin{equation}
\begin{split}
\text{CodeBLEU} = \alpha \cdot \text{Ngram} + \beta \cdot \text{Weighted} \\
+ \gamma \cdot \text{Syntax} + \delta \cdot \text{Dataflow}
\end{split}
\label{eq:codebleu}
\end{equation}
where $Ngram$ is the standard BLEU score, $Weighted$ is the weighted n-gram match based on keywords, $Syntax$ measures AST similarity, and $Dataflow$ evaluates data dependency similarity. Here, $\alpha = \beta = \gamma = \delta = 0.25$ are the default weights. The CodeBLEU scores range from 0 to 1, higher values indicating better performance.

\subsection{Membership Inference Analysis}
To evaluate the privacy risks of \llmsforcode and their compressed counterparts, we conduct a comprehensive privacy risk assessment using different MIs. MI leverages the behavioral discrepancies between a model's responses to member (training) and non-member (unseen) data to infer the presence of specific samples in the training set. This analysis is crucial for understanding the trade-off between the efficiency gains from compression and the potential increase in privacy leakage. We analyzed four distinct MI methods across various model variants (Section \ref{sec:MIAs}), and employed established metrics to quantify their effectiveness (Section \ref{sec:EvalMIA}).

\subsubsection{Methods}
\label{sec:MIAs}
MI aim to predict whether a specific data sample $x$ was included in the training dataset $D$ used to build an LLM4Code $M$. This is achieved by computing a membership score $f(x; M)$, which is then compared against a predefined threshold to infer the sample’s membership status. In our study, the model $M$ is an autoregressive language model that generates a probability distribution over the next token based on a given sequence of previous tokens, denoted as $P(x_t \mid x_1, \ldots, x_{t-1}; M)$. We explore four distinct MI methods, each utilizing a different scoring function $f$ to evaluate the likelihood of membership.

\begin{itemize}[wide=0pt]
    \item \textbf{LOSS:} An adversary leverages the model’s loss values to determine membership \cite{yeom2018privacy} using \autoref{eq:loss}. The assumption is that models tend to exhibit lower loss (or higher confidence) on samples they have seen during training compared to unseen ones. Given a threshold $K$, a sample is classified as a member if its loss falls below $K$ and as a non-member otherwise. The $k$ is determined empirically using a validation set. Following prior work, we used the area under the ROC, which is threshold-independent \cite{duan2024membership}.
    \begin{equation} 
        \label{eq:loss}
        f_{\text{loss}}(x, M) = \mathcal{L}(M, x) 
    \end{equation}
    where \( \mathcal{L} \) represents the loss function (\eg cross-entropy loss).
    
    \item \textbf{MIN\_K:} It builds on the intuition that models generalize differently for training and non-training samples (\cite{shi2023detecting}). Instead of relying solely on a single loss threshold, this method considers the $K$ smallest losses across multiple queries for a given sample. If the lowest $K$ losses are consistently smaller for certain samples, they are more likely to be members of the training dataset. This method helps mitigate variability in loss-based inference.
    \begin{equation} 
        \label{eq:min_k}
        f_{\text{min-k}}(x, M) = \min_{i \leq k} P(y_i \mid x, M)
    \end{equation}
    where \( P(y_i \mid x, M) \) represents the model’s predicted probability for class \( y_i \), and \( k \) is a tunable parameter.

    \item \textbf{ZLIB:} It utilizes data compression as an indicator of memorization \cite{carlini2021extracting}. The idea is that a model trained on a dataset tends to output more redundant and predictable sequences when queried with member samples. By compressing the model’s outputs using ZLIB, we can measure the compression ratio—samples that compress significantly more than others are likely to be members. The membership score is calculated using \autoref{eq:zlib}.
    \begin{equation}
        \label{eq:zlib}
        \frac{\log(P_v)}{\text{zlib}(x)}
    \end{equation}
    Here, $P_v$ denotes the perplexity of the model for the input sequence, and the denominator represents its ZLIB-compressed length. 
    
    \item \textbf{Reference (REF.):} This method evaluates the difference between the loss ($f_{\text{loss}}(x, M)$) of the target model and a reference model \( M_{\text{ref}} \), another LLM trained on a disjoint set of training data drawn from a similar distribution. It helps to calibrate target scores, enhancing precision and minimizing false negatives~\cite{duan2024membership}.
    \begin{equation}
        \label{eq:ref}
        f(x; M) = \Delta L(x) = L(x) - L_{\text{ref}}(x)
    \end{equation}
\end{itemize}

\subsubsection{Evaluation metrics}
\label{sec:EvalMIA}
We quantify the effectiveness of MI in distinguishing between member and non-member samples using the following standard evaluation metrics:
(1) \textbf{ROC\_AUC} (Receiver Operating Characteristic Area Under Curve) measures the ability of the MI methods to separate member and non-member samples across different decision thresholds. A higher ROC-AUC score reveals greater privacy risk. (2) \textbf{PR\_AUC} (Precision-Recall Area Under Curve) captures the trade-off between precision and recall. A high PR-AUC signifies that the MI effectively identifies training data while minimizing false positives. 

\section{Results}
In this section, we evaluate the results of the experiments to answer the research questions listed in Section \ref{sec:RQ}. Specifically, we perform our experiments of RQ1 and RQ2 based on a model family named Pythia. We choose Pythia because of its open-source availability, widely used for code completion, diverse range of model sizes, well-documentation about the member/training datasets - making it well-suited for evaluating privacy risks. \fix{1.1}{Moreover, RQ1 and RQ2 mainly focus on the impact of one variable, i.e., different precision levels, within an individual model family. Introducing multiple model families at this stage would result in a multivariable system, making the results harder to interpret and reducing the clarity of the underlying reasoning.}
In RQ3, we verify whether our findings are generalizable by considering multiple other families of \llmsforcode, such as GPT-Neo and CodeGen. 
We demonstrate our results mainly by charts for easy-to-understand purpose, and attach all the detailed numbers in our replication package.
\subsection*{\textbf{RQ1: Impact of Quantization on Privacy Risk}}\label{sec:rq1}

\begin{figure*}[ht]
    \centering
    \begin{subfigure}[b]{0.24\textwidth}
        \centering
        \includegraphics[width=\textwidth]{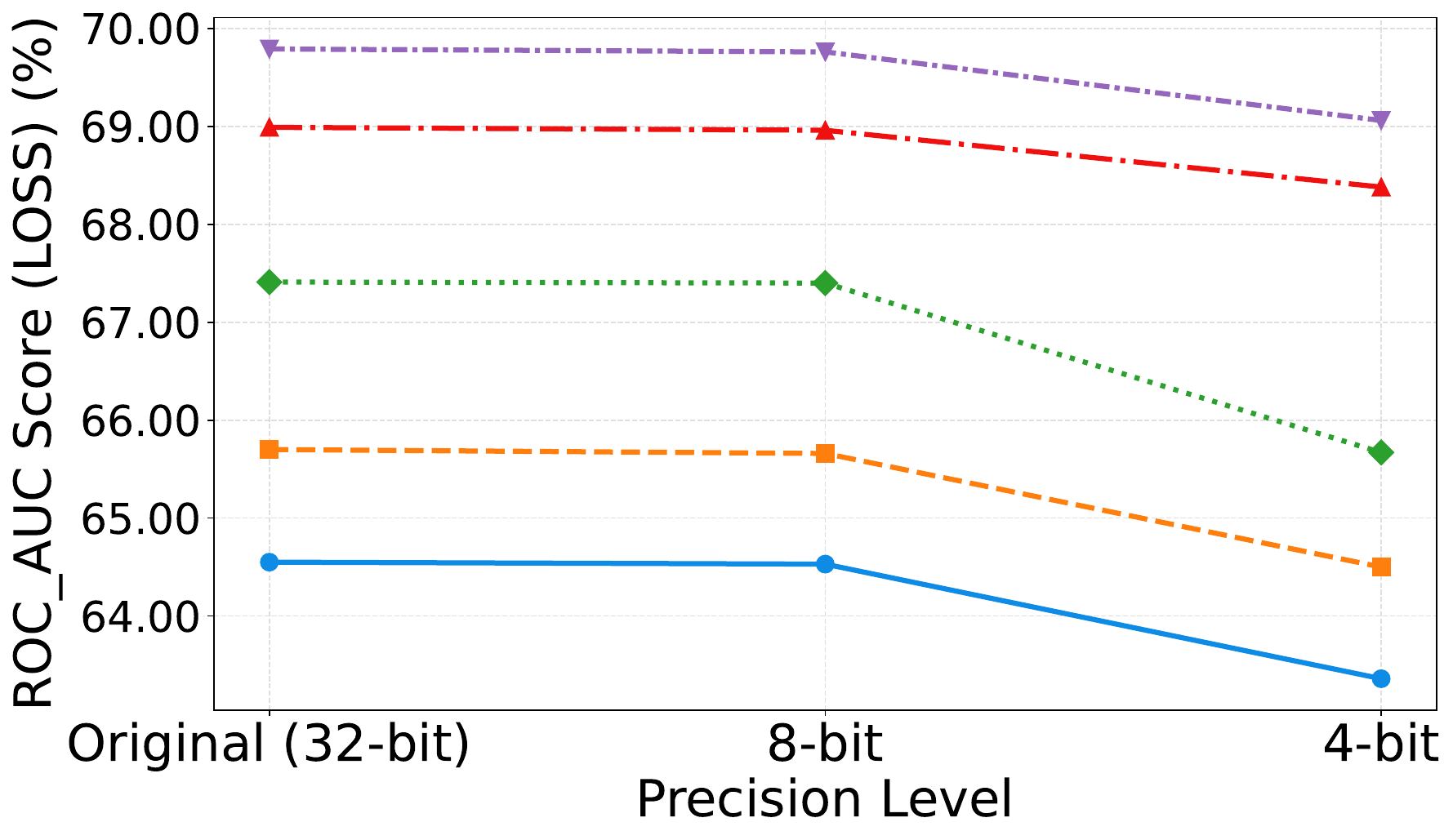}
        \caption{LOSS}
    \end{subfigure}
    \hfill
    \begin{subfigure}[b]{0.24\textwidth}
        \centering
        \includegraphics[width=\textwidth]{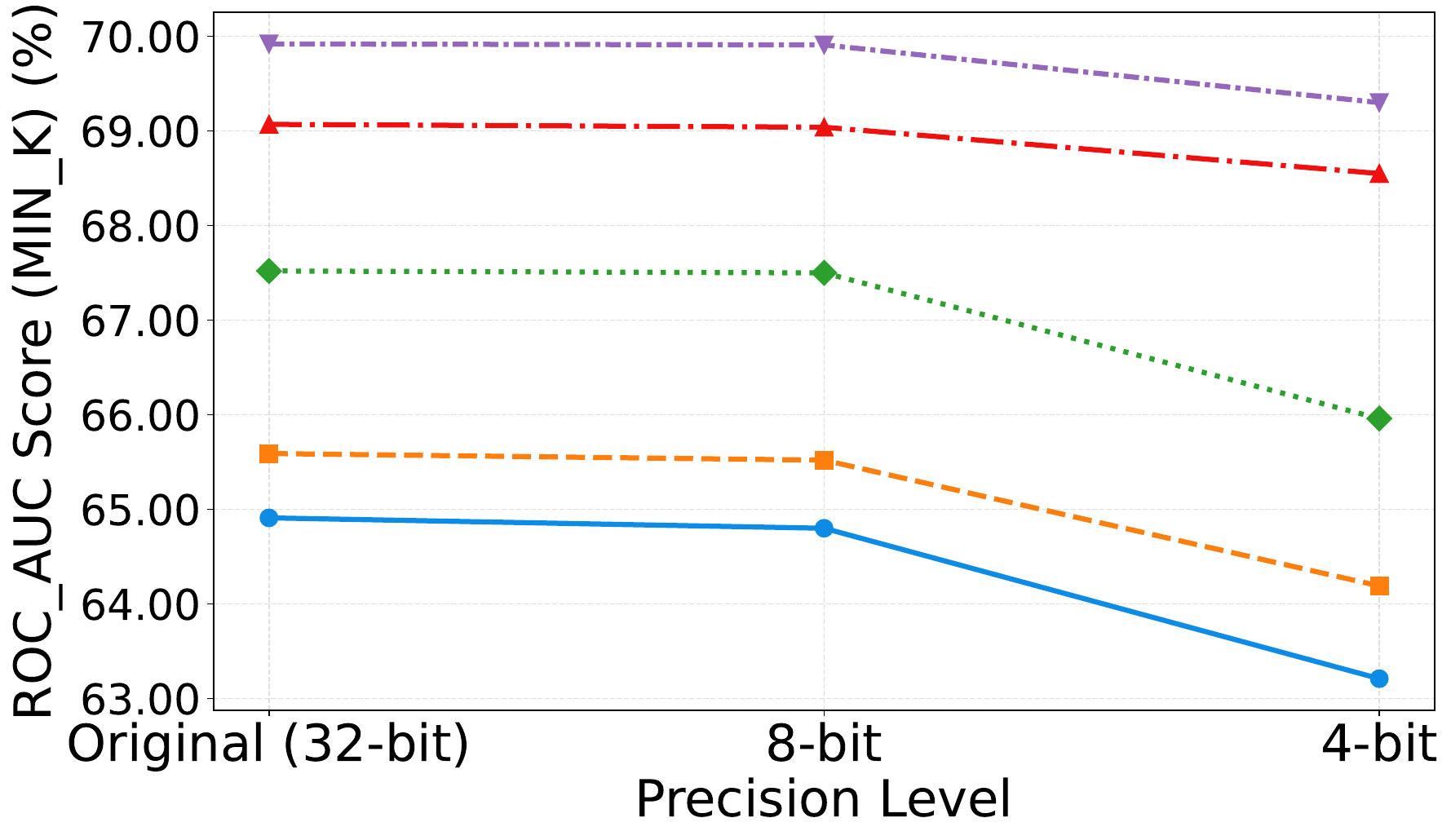}
        \caption{MIN\_K}
    \end{subfigure}
    \hfill
    \begin{subfigure}[b]{0.24\textwidth}
        \centering
        \includegraphics[width=\textwidth]{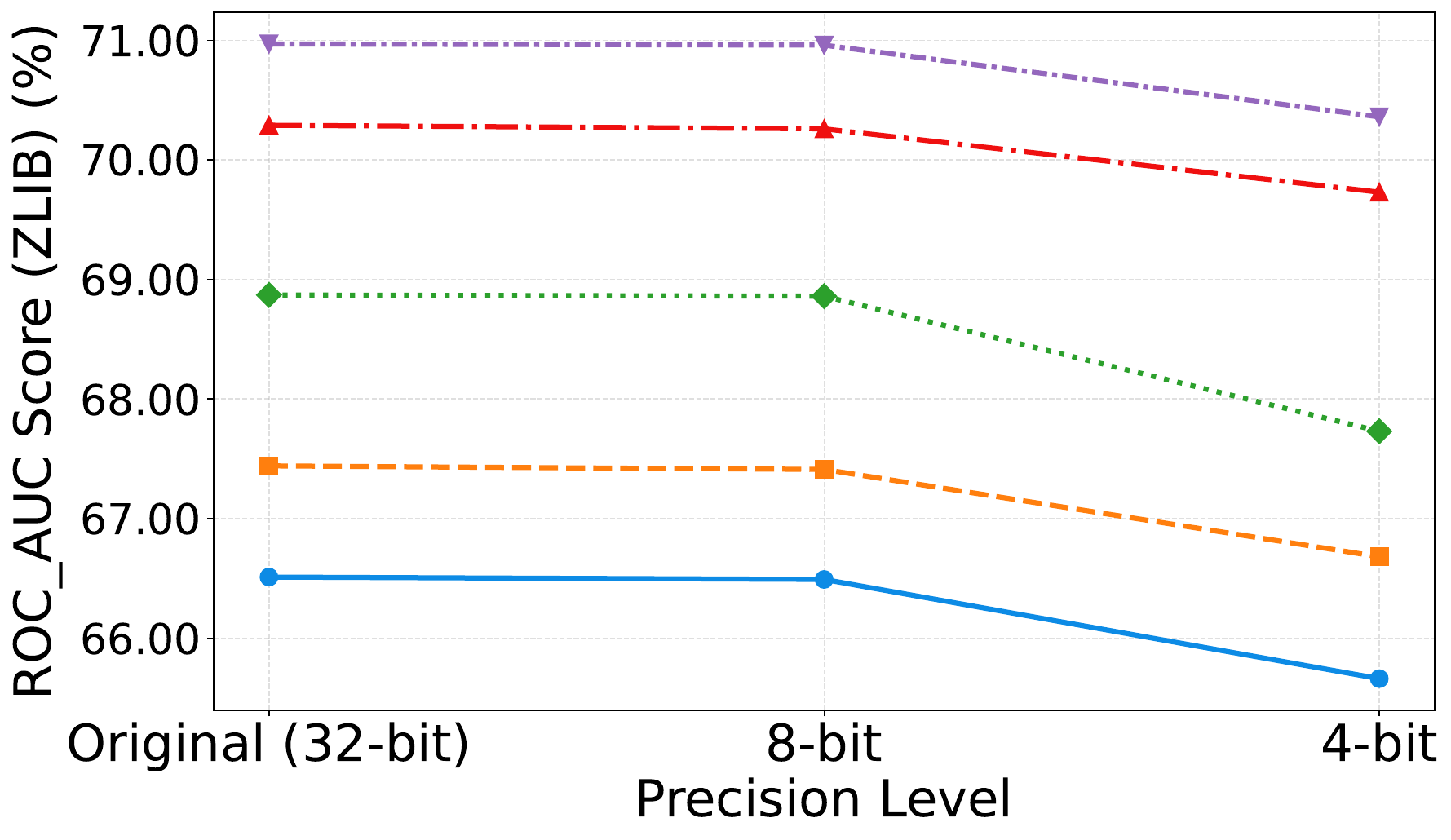}
        \caption{ZLIB}
    \end{subfigure}
    \hfill
    \begin{subfigure}[b]{0.24\textwidth}
        \centering
        \includegraphics[width=\textwidth]{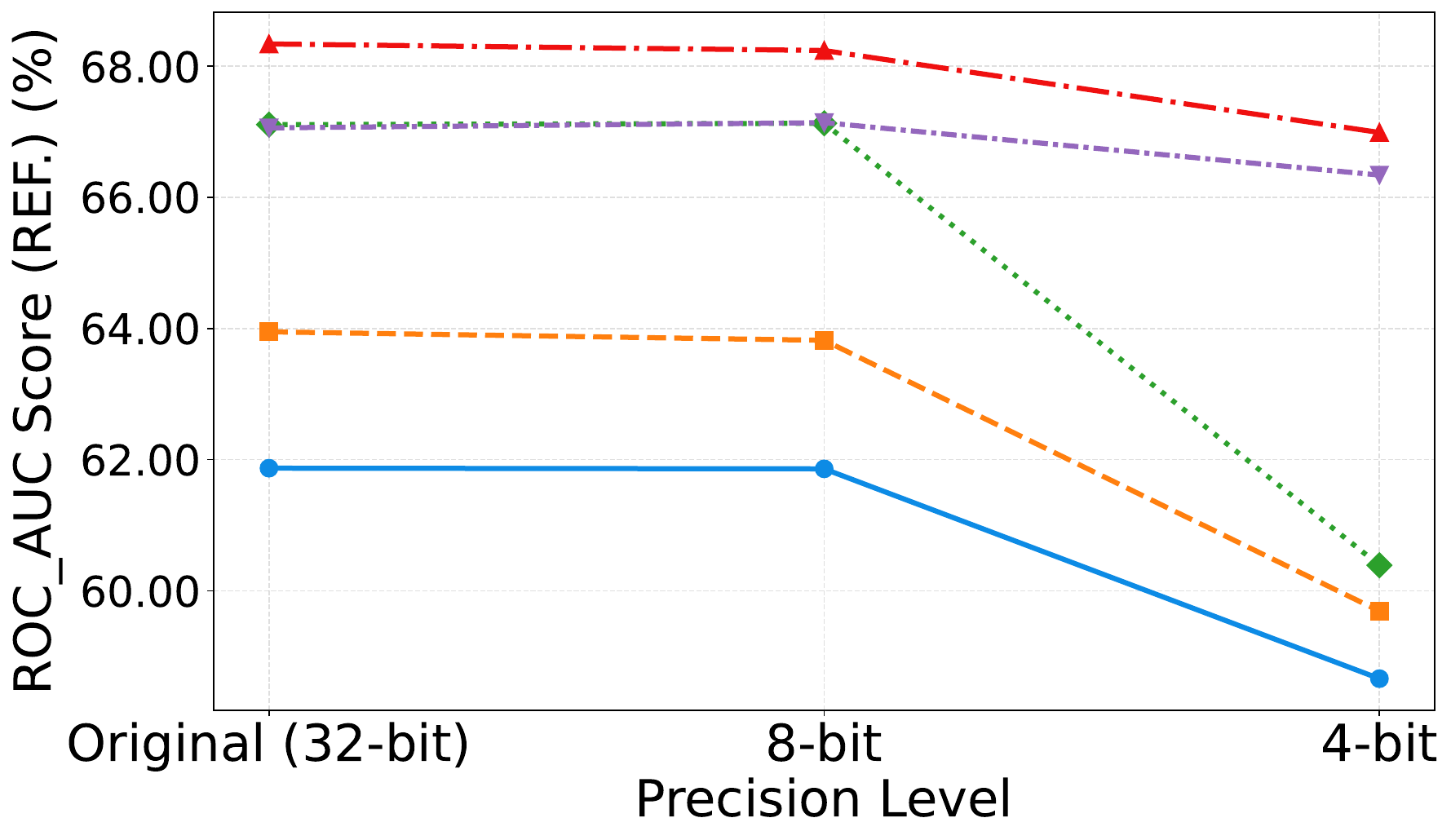}
        \caption{REF.}
    \end{subfigure}
    \vspace{-3mm}
    \caption{Comparison of privacy risks measured by MI methods in terms of \textbf{ROC\_AUC} over different precision levels of quantization. Lines represent different Pythia models distinguished by parameter sizes: pythia-70M (\textcolor[HTML]{0d8be5}{blue}), pythia-160M (\textcolor[HTML]{ff7f0e}{orange}), pythia-410M (\textcolor[HTML]{2ca02c}{green}), pythia-1B (\textcolor[HTML]{ef1010}{red}), and pythia-1.4B (\textcolor[HTML]{9467bd}{purple}).}
    \label{fig:RQ2.1}
\end{figure*}

To address RQ1, we consider four well-known MI methods—\texttt{LOSS}, \texttt{MIN\_K}, \texttt{ZLIB}, and \texttt{REF}. And we use commonly-used performance metrics of MI, i.e., ROC\_AUC and PR\_AUC (described in Section~\ref{sec:MIAs}), across five Pythia model variants. We investigate three precision levels: the Original model with full precision (i.e., 32-bit), 8-bit, and 4-bit static quantization. Here, we mainly present our results in terms of ROC\_AUC since we observed the same pattern on PR\_AUC. All the detailed results can be found at our replication package. We visualize the results in Figures \ref{fig:RQ2.1}.

\textbf{From original to 8-bit quantization.} Overall, for 8-bit quantization, we observe consistent degradation (up to 0.18\%) in MI performance across both metrics. For instance, in the \texttt{MIN\_K} method on Pythia-70M, ROC\_AUC drops from 64.91 to 64.80 (0.17\%)
. The most notable decreases appear in \texttt{REF.} method, with Pythia-160M showing a 0.20\% reduction in ROC\_AUC (63.95 → 63.82) and Pythia-1B exhibiting a 0.15\% decrease (68.34 → 68.24). Similar patterns emerge for PR\_AUC across models, with Pythia-160M decreasing by 0.12\% (58.17 → 58.10) and Pythia-1B by 0.13\% (63.68 → 63.60).

\textbf{From original to 4-bit quantization.} Similarly, 4-bit quantization degrades MI effectiveness (up to 10.01\%) across all evaluation metrics.
The Pythia-410M model shows the most substantial degradation, with ROC\_AUC for \texttt{REF.} method dropping by 10.01\% (67.11 → 60.39).
Similarly, Pythia-160M exhibits a 6.66\% reduction in \texttt{REF.} method ROC\_AUC (63.95 → 59.69).
These substantial reductions are consistent across all different model sizes, with even the largest Pythia-1.4B model showing a 2.38\% drop in \texttt{REF.} method.

\begin{figure}[h]
    \centering
    \begin{subfigure}[b]{0.4\textwidth}
        \centering
        \includegraphics[width=\textwidth]{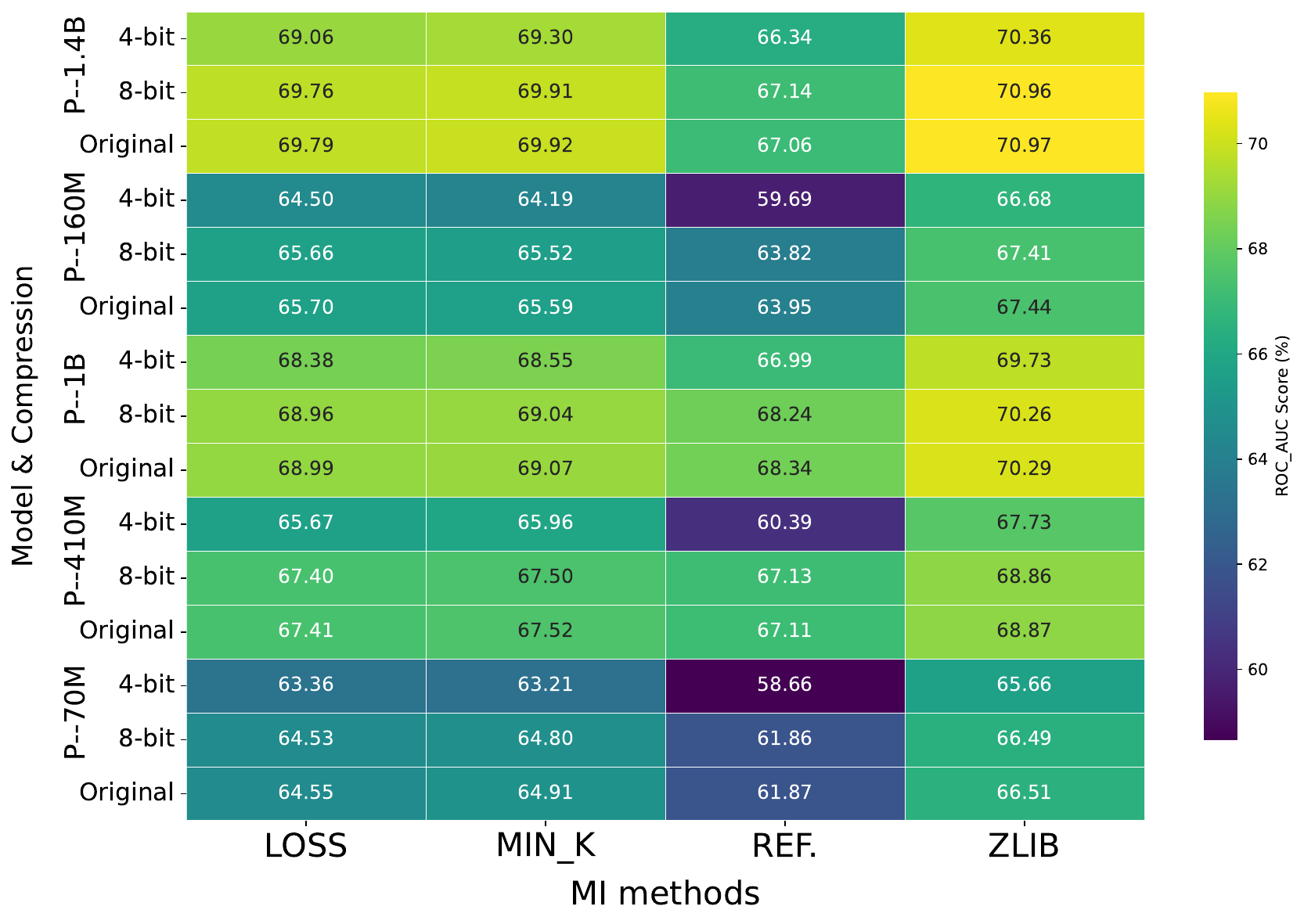}
    \end{subfigure}
    \vspace{-3mm}
    \caption{Comparison of MI effects (on ROC\_AUC) for different levels of quantization precision. The color intensity represents the magnitude of the scores, where yellow shades indicate higher privacy risk (i.e., higher AUC scores), and bluer shades indicate lower privacy risk. Each cell displays the corresponding AUC value for a specific model, quantization level, and MI methods (\texttt{LOSS, MIN\_K, REF., and ZLIB}).
    }
    \label{fig:heatmap}
    \vspace{-3mm}
\end{figure}

To better understand the susceptibility of different models to MI under varying quantization levels, we present heatmaps in \autoref{fig:heatmap} comparing ROC\_AUC scores.
These scores reflect how successfully an adversary can infer data membership, with higher values indicating greater privacy risk. The \texttt{ZLIB}-based method consistently achieves the highest AUC scores across most configurations, suggesting they are the most effective among the four methods evaluated. Additionally, we observe that the impact of quantization varies with model size: smaller and medium-sized models (e.g., pythia-70M to pythia-410M) exhibit more variation in privacy risk when quantized, whereas larger models (e.g., pythia-1B and pythia-1.4B) show relatively stable vulnerability across compression levels.

Moreover, to assess statistical significance for privacy risk, we apply the DeLong test \cite{6851192}, which is widely used to compare the differences in terms of ROC\_AUC metric.
Across all MI methods, quantization significantly reduces the privacy risk ($p<0.05$). Moreover, we observe that the p-values vary with quantization precision, 8-bit quantization yields a smaller negative p-value, whereas 4-bit quantization results in a comparatively larger negative p-value, indicating stronger statistical significance. Detailed numbers are provided in our replication package.

\begin{boxK}
\vspace{-2mm}
\textit{\textbf{RQ1:}} Both 8-bit and 4-bit static quantization significantly impacts MI effectiveness, particularly in small and medium-sized models. It indicates quantization can reduce privacy risks compared to the original models. Compare with 8-bit static quantization, 4-bit static quantization consistently reduces MI effectiveness even further.
\vspace{-2mm}
\end{boxK}
\subsection*{\textbf{RQ2: Relationship between Task Performance and Privacy Risk}}\label{sec:rq2}

\begin{figure*}[]
    \centering
    \includegraphics[width=.8\textwidth]{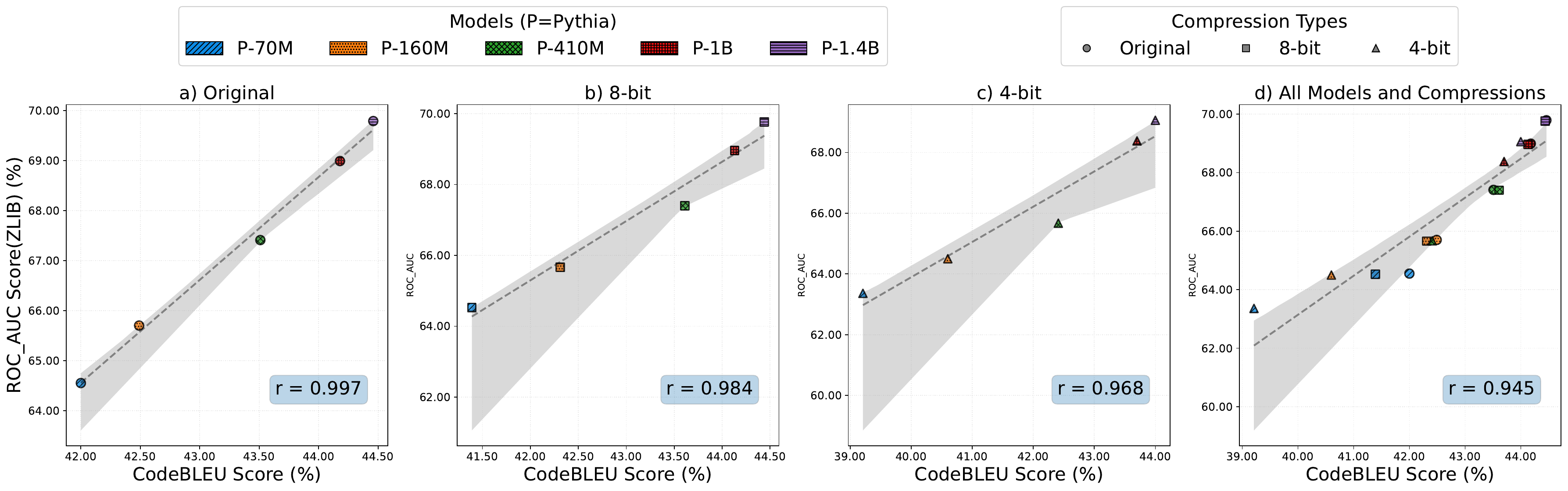}
    \vspace{-3mm}
    \caption{Correlation between task performance and MI effectiveness (\texttt{ZLIB} method) of Pythia models.
    } 
    \label{fig:3.1}
\end{figure*}

To answer RQ2, we examine the correlation between task performance (measured by CodeBLEU score) and MI performance (measured by ROC\_AUC and PR\_AUC) across all quantized variants of the Pythia models. We use a widely-used statistical metric, i.e., Pearson correlation coefficient (r)~\cite{cohen2009pearson}, to quantify the correlation relationship between task and MI performance.
It measures the strength and direction of a linear relationship between task performance and MI effectiveness, and its value ranges from -1 to 1. A value closer to 1 or -1 indicates a stronger positive or negative correlation, respectively, while a value near 0 suggests little to no linear correlation. Our results for the ZLIB method are shown in \autoref{fig:3.1}, to aid interpretation, we plotted regression lines along with shaded confidence intervals as bands. Narrower bands indicate higher certainty and consistency in the relationship. \autoref{fig:3.1}.a), \autoref{fig:3.1}.b) and \autoref{fig:3.1}.c) illustrate the correlation at each precision level: original, 8-bit, and 4-bit, respectively, where \autoref{fig:3.1}.d) represents the overall correlation that includes all precision levels to assess the overall trend. \fix{1.5}{We include the results of the other MI methods (\texttt{LOSS}, \texttt{MIN\_K}, \texttt{REF}) in our replication package since we observe the same pattern.}



\textbf{Correlation at each precision level.} Our results show that original models exhibit the strongest positive correlation between task performance and MI performance. For example, in \autoref{fig:3.1}.a) of \texttt{ZLIB} method, the Pearson correlation value of the original models is $r=0.996$ and has a narrower confidence band, suggesting stronger correlation. For 8-bit quantization, although the drop in task performance is minimal (Section \ref{sec:prelim}), we observe a statistically significant degradation in MI effectiveness (Section \ref{sec:rq1}). Hence, the relationship becomes moderately weaker ($r=0.981$) in 8-bit quantized models (\autoref{fig:3.1}. b)) and further reduced ($r=0.976$) in 4-bit quantized models (\autoref{fig:3.1}. c)), suggesting that quantization with a lower precision level dampens the correlation.

\textbf{Overall correlation.} As shown in \autoref{fig:3.1}.d), we can see that the aggregated data (all precision) still shows a positive correlation between task performance and MI effectiveness, though with more dispersion. For instance, the Pearson correlation value for the \texttt{ZLIB} method is $r=0.935$, indicating a stronger positive correlation. Additionally, a wider band of the left side of the graph indicates that 4-bit quantization of smaller models (e.g., Pythia-70M, Pythia-160M) hampers the overall correlation. 

This overall trend supports the observation that utility and privacy risk tend to increase or decrease simultaneously, but the effect may vary depending on the level of quantization applied. From the confidence interval of the regression line, it is evident that the quantized versions (8-bit, 4-bit) of smaller models (e.g., Pythia-70M, Pythia-160M) have wider bands, indicating weaker correlation, whereas original models still hold the stronger correlation. 

\begin{figure}[ht]
    \centering
    \includegraphics[width=0.4\textwidth]{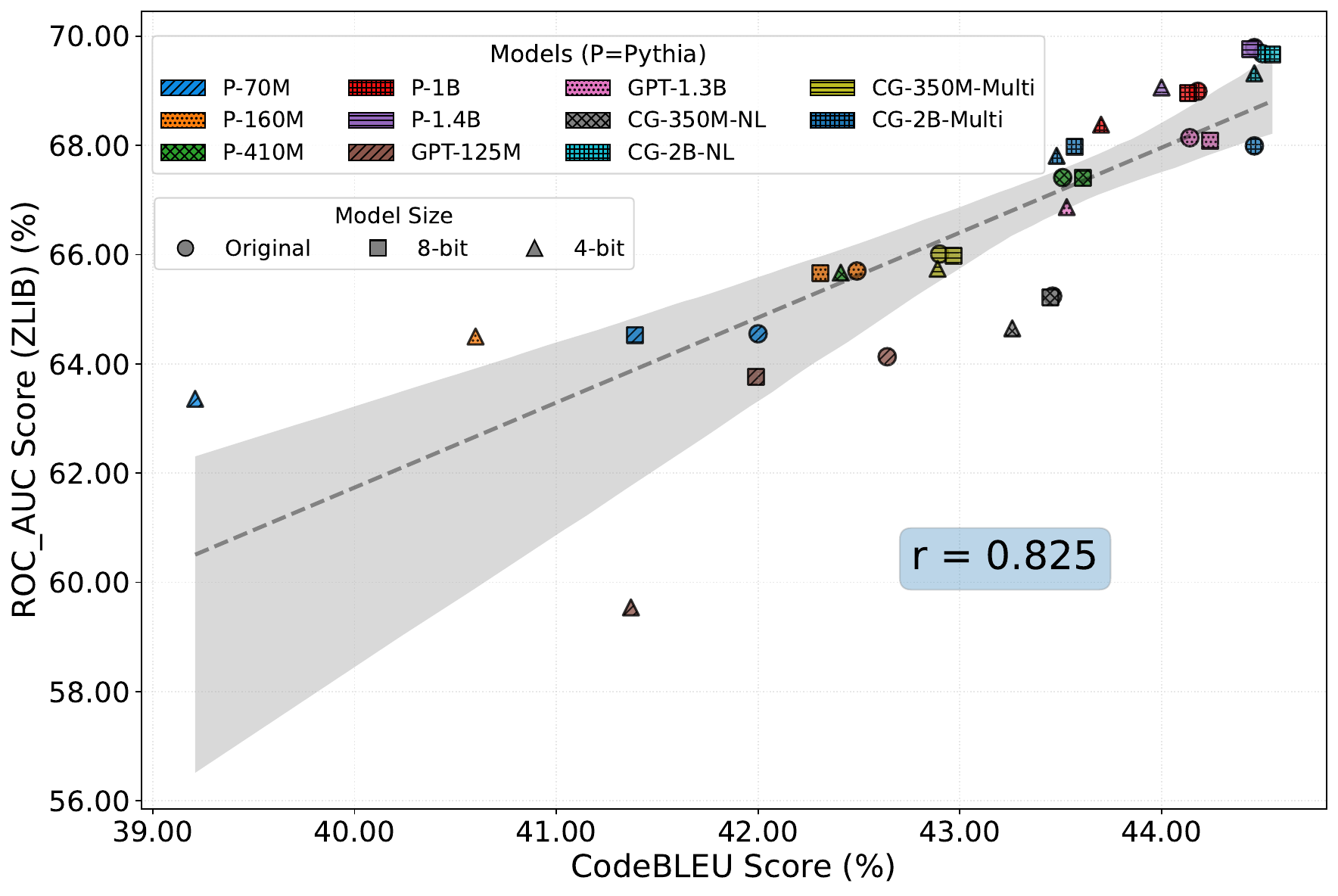}
    \vspace{-3mm}
    \caption{Correlation between task performance and MI effectiveness (\texttt{ZLIB} method) of all considered models.}
    \label{fig:4.3.3}
    \vspace{-2mm}
\end{figure}
\begin{boxK}
\vspace{-2mm}
\textit{\textbf{RQ2:}} Our results demonstrate a positive correlation between task performance and privacy risk across all MI methods. This suggests a trade-off between task performance (functional) and privacy risk (non-functional) in \llmsforcode. We also observe a stronger correlation in original models and progressively weaker ones in quantized variants, especially at 4-bit quantization - suggesting that quantization reduces the strength of this relationship.
\vspace{-2mm}
\end{boxK}

\subsection*{\textbf{RQ3: Generalizability of Our Findings}}






To answer RQ3, along with the models in Pythia family, we further consider two more popular model families, i.e., \texttt{GPT-Neo} and \texttt{CodeGen}, with multiple variants.
We perform the same steps of answering RQ1 and RQ2 on these \llmsforcode to evaluate the generalizability of our findings.

\textbf{Task performance.} 
Our results demonstrate consistent overall trends with our earlier findings, indicating that 8-bit quantization has a negligible impact on CodeBLEU scores, typically within ±1\%, thereby preserving task performance across models and scales. Also, consistently, 4-bit quantization consistently degrades CodeBLEU, with smaller models like GPT-Neo-125M and CodeGen-350M experiencing statistically significant performance drops exceeding 2–3\% ($p<0.05$). These trends mirror those observed in Pythia model family, that performance degradation due to lower-precision quantization is most pronounced in smaller-scale models.

\textbf{MI effectiveness.} We also observe a consistent pattern: 8-bit quantization has a significant impact on privacy risk as compared to original models, while 4-bit quantization degrades MI performance drastically. \fix{1.5}{We see the same pattern in other types of methods (the detailed numbers are attached in our replication package).} The relationship between task performance and privacy risk remains consistent—models with higher precision tend to be more functionally effective (better code completion). but more susceptible to MI, confirming an effectiveness-privacy leak trade-off.

\textbf{Correlation between task performance and privacy risk.} As shown in \autoref{fig:4.3.3}, we observe a similar correlation trend with the Pythia models; however, the Pearson correlation values are lower ($r \leq 0.90$). This is mainly due to the inclusion of more smaller models (e.g., GPT-NEO-125M and CodeGen-350M). These models have wider confidence bands, indicating weaker correlations.

These findings confirm that the effects of quantization on both task performance and privacy risk are model-agnostic. The patterns are stable and reproducible across model families, parameter scales, and quantization levels, supporting our conclusions' external validity and generalizability.

\begin{boxK}
\vspace{-2mm}
\textit{\textbf{RQ3:}} 8-bit quantization degrades task performance slightly but significantly reduce privacy risk drops across all model sizes and architectures, making it a safe choice for deployment. 4-bit quantization degrades both task performance (up to 7\% CodeBLEU drop in small models) and MI effectiveness (10–19\% decline), with larger models being more resilient. The effectiveness-privacy trade-off (higher CodeBLEU implies stronger MI risk) holds generally, confirming that model performance and privacy risks are intrinsically linked, regardless of the architecture.
\vspace{-2mm}
\end{boxK}

\section{Discussion}
We analyze our findings and the corresponding implications. Next, we outline potential threats to the validity of our results.

\subsection{Findings and Implications}



\begin{figure}[ht]
    \centering
    \begin{subfigure}[b]{0.4\textwidth}
        \centering
        \includegraphics[width=\textwidth]{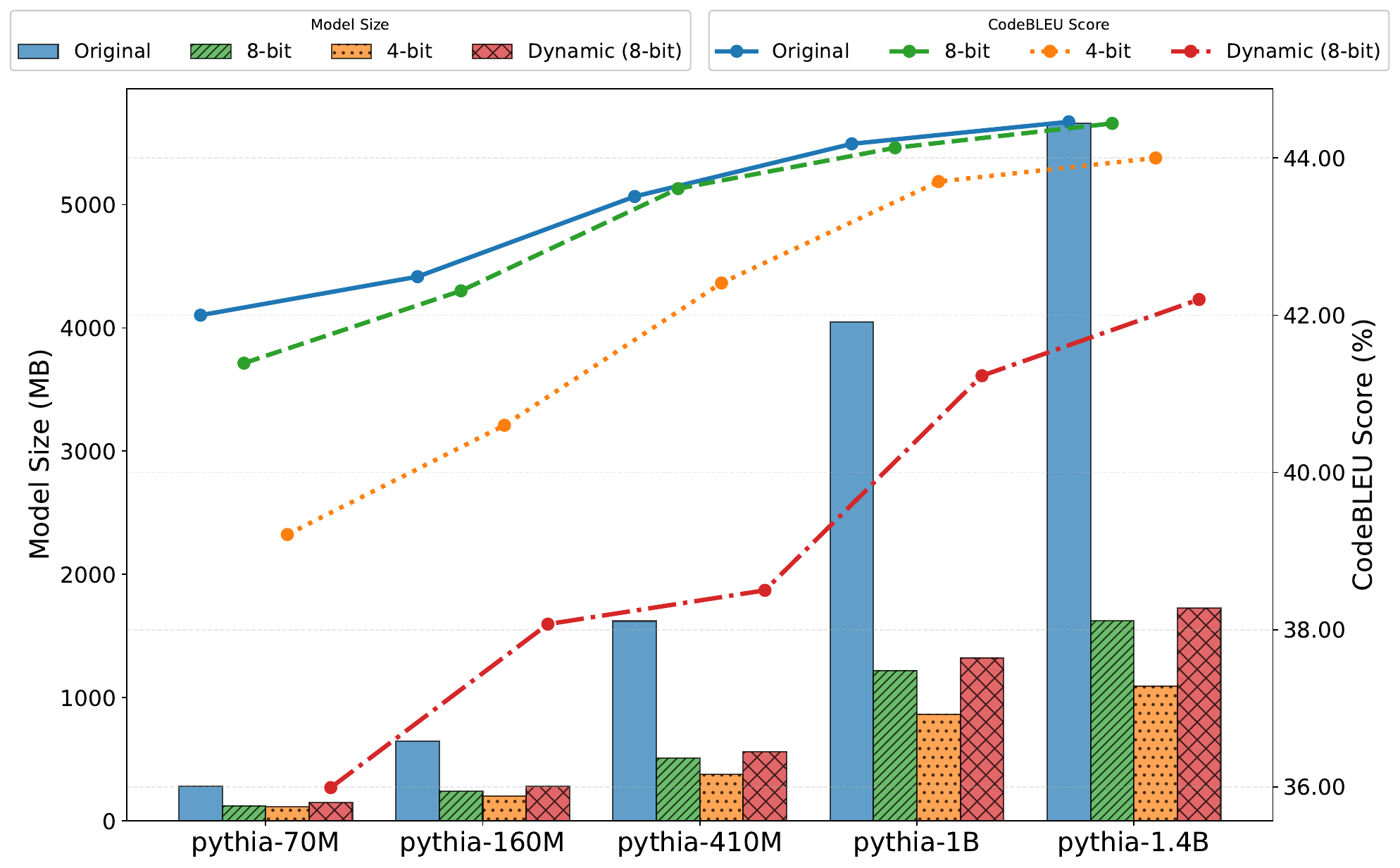}
        \label{fig:dis:task}
        \vspace{-3mm}
        \caption{Comparison of model sizes and task performance (CodeBLEU) across quantization levels.} 
    \end{subfigure}
    \hfill
    \begin{subfigure}[b]{0.4\textwidth}
        \centering
        \includegraphics[width=\textwidth]{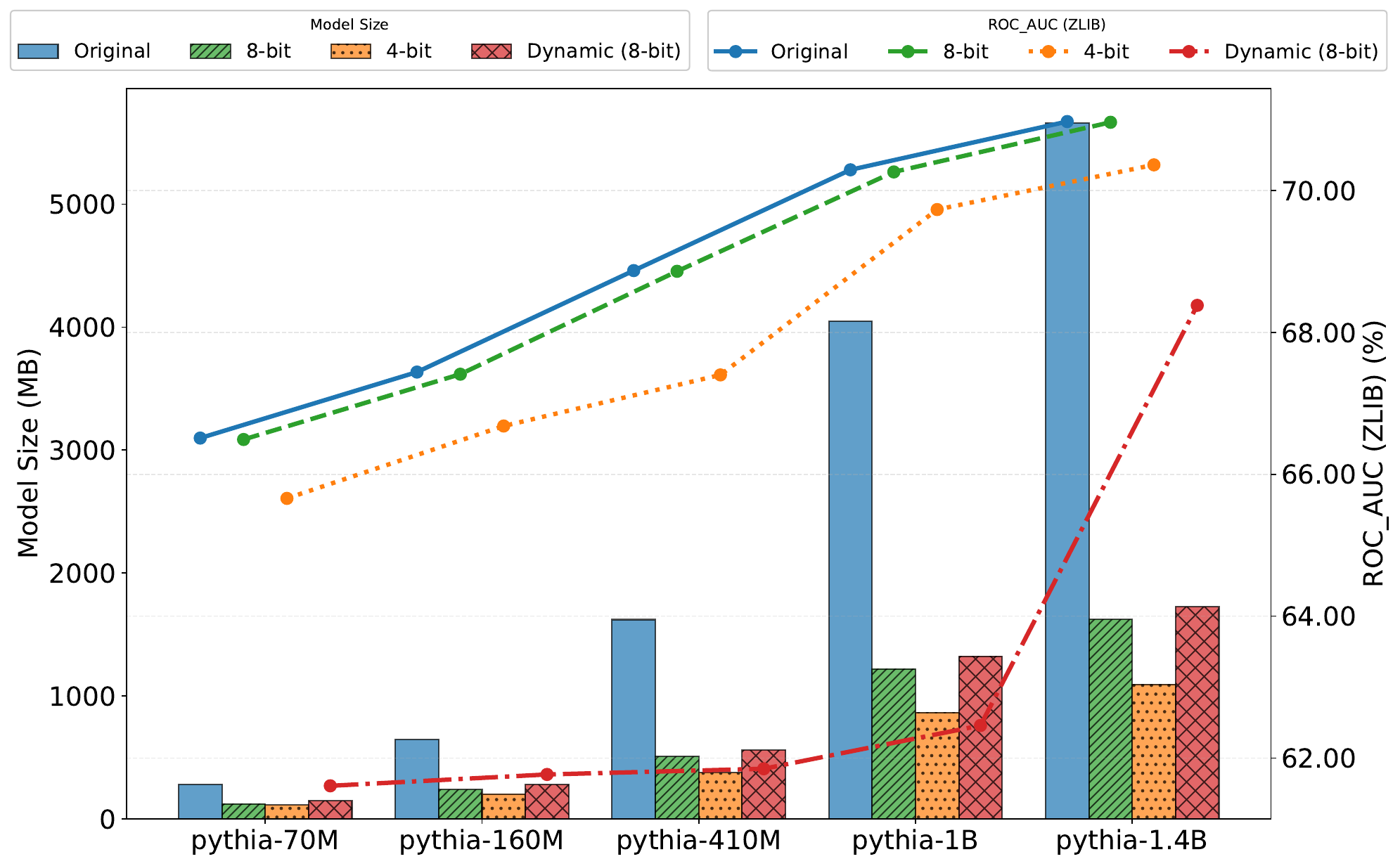}
        \label{fig:dis:mi}
        \vspace{-3mm}
        \caption{Comparison of model sizes and MI effectiveness (\texttt{ZLIB} (ROC\_AUC)) across quantization levels.} 
    \end{subfigure}
    \vspace{-3mm}
    \caption{Task performance and MI effectiveness vs model size across quantization levels for Pythia models.}
    \vspace{-3mm}
    \label{fig:dis:task-mi}
\end{figure}

\noindent\textbf{\#1 The correlation between task performance and privacy risk in LLMs4Code quantization, and the tradeoffs:}
We visualize the relationships between task performance and privacy risk in \autoref{fig:dis:task-mi}.
Specifically, for task performance (\autoref{fig:dis:task-mi}.(a)), 8-bit static quantization maintains task performance with minimal degradation (0.07\%) with respect to full precision. While 4-bit static and 8-bit dynamic quantization harm task performance, with drops of 3.15\% and 9.60\%, respectively.
Hence, from the task performance perspective, 8-bit static quantization is a more practical choice for real-world deployments.
For privacy risk (\autoref{fig:dis:task-mi}.(b)), dynamic quantization dramatically (8.8\%) reduces privacy risk, while 4-bit and 8-bit static quantization reduce by 1.1\% and 0.04\%, respectively.
Overall, our results show a positive correlation between task performance and privacy risk across different quantization methods and settings.
Our results inform model developers the impact of privacy risk beyond task performance during quantization and guide their decision-making with the tradeoffs.




However, we also observe that the correlation between task performance and privacy risk is not always linear.
As shown in \autoref{fig:4.3.3}, aggressive quantization (such as 4-bit static quantization) of smaller \llmsforcode (P-70M, P-160M, and GPT-125M), particularly outside the shaded region, weakens the correlation. In these cases, quantization disproportionately harms task performance compared to its modest privacy gains, weakening the performance-privacy correlation. For instance, in the case of the Pythia 70M model, task performance is approximately $2.5$ times more sensitive to quantization than MI effectiveness. CodeBLEU drops by an average of 6.6\% while ROC\_AUC drops by only 1.6\%.

\vspace{2mm}
\noindent\textbf{\#2: Quantized larger models could outperform smaller full-precision models in terms of both task performance and privacy risk:}
We investigated whether a larger model quantized with a low precision level can serve as an effective alternative to a smaller model with higher precision. Our results demonstrate that this effect is particularly pronounced when comparing quantized variants of mid-sized models, where, after quantization, bigger models matches or even smaller in size than the smaller models of full precision, but give better task performance and lower privacy risks, shown in \autoref{fig:dis:task-mi}.
For example, we can see that the model size of Pythia-410M (4-bit) is 42\% smaller (649.34 vs 376.75) and 0.06\% lower privacy risk than the Pythia-160M (full precision) model (67.44 vs 67.40), but retains 99.8\% of the original 160M’s CodeBLEU score (42.49 vs 42.41). 
This phenomenon suggests the presence of a better balance of the task performance-privacy risk trade-offs in \llmsforcode quantization when given multiple original models. We advocate for future studies to explore an automated method for efficiently identifying the optimal balance.

\vspace{2mm}
\noindent\textbf{\#3  Quantization without considering model size may limit overall gain in a resource-constrained environment:} While 8-bit quantization slightly impacts task performance and has a significant impact on privacy risk, 4-bit quantization has a significant impact on both of them. This suggests that higher-precision quantization (such as 8-bit) may be a favorable choice for deployment when preserving functional performance is the top priority. However, we also find there are cases where selecting 4-bit quantization models of larger \llmsforcode gives better task performance, reduces model size, and lowers privacy risks than the 8-bit quantization models, even full precision smaller models. This suggests that for deployments in a resource-constrained environment where resource consumption, performance, and privacy are required, lower bit quantization of larger models should also be considered.

\vspace{2mm}
\noindent\textbf{\#4 ZLIB-Based MI method remain most effective for \llmsforcode privacy risk assessment:} Across all model sizes and compression levels, ZLIB-based MI method consistently achieved the highest ROC\_AUC and PR\_AUC scores. This indicates that compression artifacts retain rich MI that can be exploited, even in heavily quantized models. The robustness of ZLIB-based MI methods underscores their practical effectiveness and highlights the need for compression-aware privacy defenses. Therefore, based on our results, we advocate that future works should consider ZLIB-based MI method to assess privacy risk for \llmsforcode.


\subsection{Threats to Validity}

Threats to \textbf{internal validity} pertain to the accuracy and reliability of our experimental implementation. To mitigate these threats, we employed the official releases of the models from HuggingFace. For quantization techniques, we utilized implementations by PyTorch libraries, ensuring methodological correctness. Additionally, we reused the original implementations of four MI methods provided by the authors of~\cite{duan2024membership}
, and followed their configurations.
Threats to \textbf{external validity} concern the generalizability of our findings across different models and settings.
One potential threat lies in model architecture; our findings may not extend beyond the specific architectures studied. However, our focus on causal language models—namely Pythia, CodeGen, and GPT-Neo—provides a strong foundation, as GPT-based models underpin many state-of-the-art systems, and CodeGen has demonstrated strong performance in code generation. These choices enhance the external validity and relevance of our findings.
Threats to \textbf{construct validity} relate to the appropriateness of the metrics and evaluation procedures used. We assessed model effectiveness and MI performance,
we employed widely used metrics in the literature~\cite{duan2024membership, zhou2024gotcha}. Therefore, we consider the threat to construct validity to be minimal.

\section{Conclusion}
In this paper, we explored the impact of \llmsforcode quantization combines both functional (task performance) and non-functional (privacy risk) aspects. Our study across three model families reveals that 8-bit static quantization slightly impacts task performance, while showing a significant drop in privacy risk compared to original models. However, 4-bit static and 8-bit dynamic quantization introduce a clearer privacy benefit by lowering MI effectiveness, but at the cost of decreased task performance. These results highlight a nuanced effectiveness-privacy trade-off.
We also uncover that aggressively quantized larger models could outperform full-precision smaller models on both aspects. This suggests the potential of strategic compression as a dual-purpose tool for \llmsforcode quantization.
Future work may investigate adaptive or hybrid quantization strategies that identify the optimal balance of the tradeoffs. 

\bibliographystyle{ACM-Reference-Format}
\bibliography{ref}
\end{document}